\documentclass[numbib]{imamat}

\usepackage{graphicx} 

\usepackage[usenames,dvipsnames]{color}

\newcommand{\mylab}[1]{\label{#1}}
\renewcommand{\vec}[1]{\mathbf{#1}}
\newcommand{\vecg}[1]{\boldsymbol{#1}}
\newcommand{\tens}[1]{\mathbf{\underline{#1}}}

\begin{document}

\title{Localized states in coupled Cahn-Hilliard equations}

\shorttitle{Localized states in coupled CH equations} 
\shortauthorlist{T. Frohoff-H\"ulsmann, U. Thiele} 
\author{
\name{Tobias Frohoff-H\"ulsmann \thanks{ORCID ID: 0000-0002-5589-9397}}
\address{Institut f\"ur Theoretische Physik, Westf\"alische Wilhelms-Universit\"at M\"unster, Wilhelm-Klemm-Str.\ 9, 48149 M\"unster, Germany
\email{Corresponding author: t\_froh01@uni-muenster.de}} 
\name{Uwe Thiele\thanks{homepage: http://www.uwethiele.de; ORCID ID: 0000-0001-7989-9271}}
\address{Institut f\"ur Theoretische Physik, Westf\"alische Wilhelms-Universit\"at M\"unster, Wilhelm-Klemm-Str.\ 9, 48149 M\"unster, Germany~\\
  Center for Nonlinear Science (CeNoS), Westf{\"a}lische Wilhelms-Universit\"at M\"unster, Corrensstr.\ 2, 48149 M\"unster, Germany
  \email{u.thiele@uni-muenster.de}
}
}

\maketitle

\begin{abstract}
{The classical Cahn-Hilliard (CH) equation corresponds to a gradient dynamics model that describes phase decomposition in a binary mixture. In the spinodal region, an initially homogeneous state spontaneously decomposes via a large-scale instability into drop, hole or labyrinthine concentration patterns of a typical structure length followed by a continuously ongoing coarsening process. Here we consider the coupled CH dynamics of two concentration fields and show that nonreciprocal (or active, or nonvariational) coupling may induce a small-scale (Turing) instability. At the corresponding primary bifurcation a branch of periodically patterned steady states emerges. Furthermore, there exist localized states that consist of patterned patches coexisting with a homogeneous background. The branches of steady parity-symmetric and parity-asymmetric localized states form a slanted homoclinic snaking structure typical for systems with a conservation law. In contrast to snaking structures in systems with gradient dynamics, here, Hopf instabilities occur at sufficiently large activity which result in oscillating and traveling localized patterns.
	~\\
	The published version of this preprint can be found under \newline
	T. Frohoff-H\"ulsmann and U. Thiele.
	Localized states in coupled Cahn--Hilliard equations.
	\textit{IMA J. Appl. Math.} 86:924--943, 2021.
	DOI: 10.1093/imamat/hxab026
}
{Cahn-Hilliard models, localized states, slanted homoclinic snaking, conservation laws, Turing instability, nonreciprocal interaction}
\end{abstract}

\section{Introduction}
Homogeneous mixtures of different materials are often unstable and undergo phase separation (demixing or decomposition). Such processes are abundant in nature and technology where they occur for many multi-component materials ranging from solid alloys via soft gels and polymers to liquid mixtures \citep{Lang1992,Bray1994ap,GPDM2003ap,Onuki2002}. Quenching a homogeneous mixture into a parameter region where the uniform state is linearly unstable,  phase separation occurs. In the initial stage, the mixture separates on a diffusive length scale into regions of one phase surrounded by the other phase, i.e., one obtains small clusters, drops or islands. A labyrinthine structure may also emerge. Subsequently, coarsening sets in, i.e., the mean structure size continuously increases in time until ultimately a single domain-spanning structure prevails that corresponds to the global energy minimum \citep{Lang1992,Bray1994ap}.
The described behavior is typical for binary systems as well as for systems consisting of more components. The formation of stable patterns is normally not reported for such systems, in particular, they are not known for the occurrence of localized states that consist of patterned patches coexisting with a homogeneous background. However, here we show that steady parity-symmetric and parity-asymmetric localized states can occur and show homoclinic snaking \citep{WoCh1999pd,BuKn2006pre} if phase separation in a multicomponent system is considered in an out-of-equilibrium setting typical for active media \citep{FrWT2021pre}.

The common model for phase separation was introduced by Cahn and Hilliard \citep{CaHi1958jcp, Cahn1965jcp} and is widely studied as the Cahn-Hilliard (CH) equation \citep{Novi1985jsp,Lang1992,Bray1994ap,Onuki2002,Doi2013,TFEK2019njp}. In its classic form it can be presented as a mass-conserving gradient dynamics for a scalar order parameter field $\phi$ that corresponds to a scaled and shifted density or concentration\citep{Cahn1965jcp,Lang1992},
\begin{equation}
\partial_t\,\phi\,=\,\vec{\nabla}\cdot \left[M \vec{\nabla}\,
\frac{\delta F[\phi]}{\delta \phi}\right].
\mylab{eq:cht}
\end{equation}
Here, $M$ is a positive mobility constant and the underlying free energy functional is \citep{CaHi1958jcp}
\begin{equation}
F[\phi(\vec{x},t)] =\int_V \left[\frac{\kappa}{2}\, |\nabla \phi|^2  + f(\phi) \right] \mathrm{d}^nr,
\mylab{eq:helmhet2}
\end{equation}
with $V$ being the considered volume. The first term in (\ref{eq:helmhet2}) penalizes interfaces ($\kappa\ge0$) and $f(\phi)$ is a simple double-well potential that is specified later. Mobility functions $M(\phi)$ may also be considered \citep{KoOt2002cmp}.

It is notable, that its character as a structure-forming but not pattern-forming model resulted in the remarkable feat that the CH equation avoided being mentioned in the monumental seminal review by Cross and Hohenberg on out-of-equilibrium pattern formation \citep{CrHo1993rmp}. In consequence, it has been slow to claim its place in the pattern formation community, in contrast to its prominence in soft matter and material science. 
However, it still made the classification of Halperin and Hohenberg \citep{HoHa1977rmp} where it is known as ``model~B''.

Nevertheless, it is known for some time that the CH equation can show more intricate behavior if model amendments account for out-of-equilibrium driving forces. One example are laterally driven phase-separating systems that are described by the convective CH equation \citep{EmBr1996pre,GoDN1998pd,GNDZ2001prl}. It corresponds to a CH equation with an additional Burgers-type convective term that breaks the gradient dynamics structure but is still mass-conserving. A recent numerical bifurcation study analyzes in detail how coarsening is suppressed by the lateral driving \citep{TALT2020n}.

The convective CH equation models phase separation and coarsening dynamics in an externally imposed gradient or flow as encountered in various processing technologies. Prominent are coating technologies like dip coating and Langmuir-Blodgett transfer that may be described by models of driven CH type (e.g., \citet{TWGT2019prf} and \citet{KoTh2014n}, respectively). However, phase separation is now also recognized as a widespread mechanism employed by biological cells for its inner organization \citep{BeBH2018rpp}. Then, the out-of-equilibrium character of living cells may crucially amend main features of phase-separation, e.g., by coupling it to chemical reactions. Such processes are studied for a number of particular systems involving membranes and/or centrosomes. For instance, John and B\"ar investigate the evolution of lipid domains in the cell membrane with a four-field model consisting of reaction-diffusion equations for proteins and regulating enzymes coupled to a CH-type equation for the lipid distributions (model~II in \citet{JoBa2005prl}). The resulting model can show slowed-down coarsening or the emergence of traveling waves instead of coarsening. Other examples are models for phase separation in ternary mixtures with chemical reactions \citep{ToYa2002jcp,OkOh2003pre}. An active emulsion model for centrosome dynamics uses a reactive coupling of two CH equations \citep{ZwHJ2015pre,WZJL2019rpp}. In a related conceptual model, two CH equations are coupled by a reactive coupling of different type \citep{SATB2014c}. There, arrest of coarsening and traveling patterns 
are described.

Beside its gradient dynamics structure, the other characteristic feature of the CH model is the mass-conserving character of the dynamics. The important role of conservation laws in pattern forming systems is increasingly recognized \citep{MaCo2000non,Cox2004pla,WiMC2005n,Knob2016jam} but their role in many specific systems still necessitates a deeper understanding. The mentioned nonvariational systems of two coupled CH equations keep either none \citep{SATB2014c} or only one \citep{ZwHJ2015pre} of the two conservation laws of the original uncoupled equations. This may partly explain the rather different reported behavior. However, no comparative study is available.\footnote{Also multi-species reaction-diffusion systems with one or two conservation laws are discussed by a number of groups. For a brief overview see the introduction of \citet{FrWT2021pre}.} 

In contrast, we consider a system of two CH equations with symmetric and antisymmetric
coupling that both preserve the full mass conservation properties of the uncoupled equations.  Sometimes, these couplings are referred to as ``variational'' and ``nonvariational'', respectively. However, this may be misleading as explained below in Section~\ref{sec:model}.
A similar system has very recently been considered by \citet{SaAG2020} and \citet{YoBM2020pnas} for a restricted (non-generic) choice of parameters. Both studies focus on the transition from phase separation into static domain-spanning clusters to self-propelled clusters caused by the antisymmetric, i.e., nonvariational coupling. It is sometimes called nonreciprocal coupling as it breaks Newton's third law \citep{IBHD2015prx}. For the general case, a further transition from phase separation into domain-spanning clusters to the emergence of stable steady patterns of finite typical length scales occurs. \citet{FrWT2021pre} analyze the associated intricate linear behavior and shows that it is closely related to the classical Turing system of two coupled reaction-diffusion systems \citep{Turi1952ptrslsbs}. Furthermore, they show that coarsening may be arrested or fully suppressed, and scrutinize the related bifurcation behavior.

Based on the results of \citet{FrWT2021pre} on the occurrence of a small-scale (i.e., Turing) instability in the two-field CH system with dominant antisymmetric coupling, 
here, we pursue the question whether, in consequence, homoclinic snaking of localized states \citep{WoCh1999pd,BuKn2006pre} occurs for this system and how it may differ from snaking behavior in other related systems. In particular, we numerically show that branches of symmetric and asymmetric localized states exist that form a slanted snakes-and-ladders structure typical for systems with a conservation law \citep{Knob2016jam,TARG2013pre}. As the nonvariational aspect of the model is essential we also expect, time-periodic localized states to occur, similar to states described for nonvariational Swift-Hohenberg equations \citep{HoKn2011pre,BuDa2012sjads} and active phase field crystal models \citep{OKGT2020c,HAGK2020arxiv}. Our analysis of the bifurcation structures is based on numerical continuation methods \citep{AllgowerGeorg1987,KrauskopfOsingaGalan-Vioque2007,EGUW2019springer} implemented in the package pde2path \citep{UeWR2014nmma,Ueck2019ccp}.

Our work is structured as follows. Section~\ref{sec:model} introduces the model, sketches the numerical method and defines the employed solution measures. The subsequent Section~\ref{sec:linstab} sketches the temporal and spatial linear stability analyses of uniform steady states and discusses under which conditions the system can show a small-scale stationary instability, i.e., a Turing instability. Section~\ref{sec:snake} then numerically analyzes the bifurcation behavior focusing on the occurrence of localized states and time-periodic behavior within the snakes-and-ladders structures. Finally, a conclusion is given in Section~\ref{sec:conc}.
%

%
%
\section{Model}
\mylab{sec:model}
The model consists of two linearly coupled CH equations in one spatial dimension. 
After nondimensionalisation the model reads
\begin{alignat}{1}
\begin{aligned}
\qquad \frac{\partial\phi_1}{\partial t}  &= \frac{1}{\ell^2}\frac{\partial^2}{\partial x^2} \left(- \frac{1}{\ell^2}\frac{\partial^2\phi_1 }{\partial x^2} +f_1'(\phi_1) - \left(\rho + \alpha\right) \phi_2 \right)\\
\qquad \frac{\partial\phi_2}{\partial t}  &=  \frac{Q}{\ell^2}\frac{\partial^2}{\partial x^2} \left(-\frac{\kappa}{\ell^2}\frac{\partial^2 \phi_2}{\partial x^2} +f_2'(\phi_2) - \left(\rho - \alpha\right) \phi_1 \right)\,
\end{aligned}
\label{eq:nondim_bG_final}
\end{alignat}
with the double-well potentials
\begin{alignat}{1}
\begin{aligned}
f_1=&  \frac{a}{2} \phi_1^2  + \frac{1}{4} \phi_1^4 \, ,\\
  f_2=& \frac{a+a_\Delta}{2} \phi_2^2  + \frac{1}{4} \phi_2^4\,.
\end{aligned}
\label{eq:ff}
\end{alignat}
As the dynamics of both fields is mass conserving, at all times
\begin{alignat}{1}
\begin{aligned}
 \int \textrm{d}x \,\phi_1 = &\phi_{0}\\
  \int \textrm{d}x \,\phi_2 =& \phi_{0} + \Delta \phi_0\,,
\end{aligned}
\label{eq:int}
\end{alignat}
where for simplicity we fix $\Delta \phi_0=0$ throughout the present work. Hence, $\phi_0$ is the mean concentration of both fields. In the following we employ $\phi_0$ as our main control parameter.

The single CH equation for $\phi_1$ without coupling has a critical point at $a=0$ (and $\phi_0=0$), i.e., decomposition can only occur for $a<0$. Since the parameter $a$ represents an effective temperature, $a_\Delta$ is the difference in critical temperatures of the uncoupled fields. Furthermore, $\kappa$ is the relative rigidity of the uncoupled fields, $\ell$ is the effective domain size. For details of the nondimensionalisation and the effective parameters see \citet{FrWT2021pre}.
The linear coupling is composed of a symmetric 
part of strength $\rho$ and an antisymmetric (nonreciprocal) 
part of strength $\alpha$. 
The model \eqref{eq:nondim_bG_final} is also called the nonreciprocal CH model.

Without the antisymmetric coupling, the model represents a gradient dynamics of form \eqref{eq:cht} and the two species attract [repel] each other for $\rho>0$ [$\rho<0$]. 
It is less obvious that the gradient dynamics character actually still remains for nonzero $\alpha$ if $|\rho|>|\alpha|$. If we rewrite Eqs.~\eqref{eq:nondim_bG_final} as
\begin{alignat}{1}
\begin{aligned}
\qquad \frac{\partial\phi_1}{\partial t}  &=   \frac{\left(\rho + \alpha\right)}{\ell^2}\frac{\partial^2}{\partial x^2} \left(- \frac{1}{  \left(\rho + \alpha\right)\ell^2}\frac{\partial^2\phi_1 }{\partial x^2} +\frac{f_1'(\phi_1)}{ \left(\rho + \alpha\right)} -  \phi_2 \right)\\
\qquad \frac{\partial\phi_2}{\partial t}  &=  \frac{Q\left(\rho - \alpha\right) }{ \ell^2}\frac{\partial^2}{\partial x^2} \left(-\frac{\kappa}{\left(\rho - \alpha\right) \ell^2}\frac{\partial^2 \phi_2}{\partial x^2} + \frac{f_2'(\phi_2)}{\left(\rho - \alpha\right) } - \phi_1 \right)\,
\end{aligned}
\end{alignat}
the form \eqref{eq:cht} is reproduced with
\begin{align}
F= & \int {\rm d} x \frac{1}{ 2 \left(\rho + \alpha\right)\ell^2} \left(\partial_x \phi_1\right)^2 + \frac{\kappa}{ 2 \left(\rho - \alpha\right)\ell^2} \left(\partial_x \phi_2\right)^2 
+\frac{f_1(\phi_1)}{ \left(\rho + \alpha\right)} +\frac{f_2(\phi_2)}{ \left(\rho - \alpha\right)} - \phi_1\phi_2~\\
\text{and} \quad M= & \left(\begin{array}{c c}
\frac{\left(\rho + \alpha\right)}{\ell^2} & 0~\\
0 & \frac{Q \left(\rho - \alpha\right)}{\ell^2}
\end{array}
\right)
\end{align}
as then the mobility matrix is still positive definite.\footnote{Here we have assumed that $\rho>0$. If $\rho<0$ one could factorize an additional sign in both equations which then leads to the same result.} Furthermore, all  interface contributions of the free energy are positive. 
However, for a dominant nonreciprocal interaction ($|\rho|<|\alpha|$), the mobility becomes negative definite and the gradient dynamics structure is broken. The interaction then represents a nonvariational contribution rendering the system active.

We emphasize that the chosen coupling ensures that both conservation laws remain valid. This implies, that it can not represent a chemical reaction between the components, but could still model the effective interaction of two conserved catalysts whose nonreciprocal coupling is mediated by reactions with other species that are not explicitly modeled \citep{CWXJ2014nb}. It could also be seen as a continuous predator-prey-type model of two potentially 'phase separating' species: e.g., for $|\alpha|>|\rho|$ species one would be attracted by species two while species two is repelled by species one, not unlike the attraction-repulsion scheme in \citet{ChKo2014jrsi}. Models similar to the one studied here are investigated by \citet{SaAG2020} and \citet{YoBM2020pnas}. However, there, non-generic cases are considered corresponding in our scaling to $Q=\kappa=1$ \citep{SaAG2020} and $\kappa=0$ \citep{YoBM2020pnas}, respectively. These small but significant differences to the general model suppress the Turing instability as discussed in the linear stability analysis in Section~\ref{sec:linstab}.
In particular $Q\kappa \neq \{0,1\} $ is a necessary condition for the existence of stable steady patterns and the related suppression of coarsening as explained by \citet{FrWT2021pre}. It is as well a condition for the existence of localized states.

In Section~\ref{sec:snake} we provide a fully nonlinear analysis of localized steady states in the form of a numerical bifurcation analysis. In particular, we study solutions of \eqref{eq:nondim_bG_final} with $\partial_t\phi_1=\partial_t\phi_2=0$ applying pseudo-arclength continuation \citep{AllgowerGeorg1987,KrauskopfOsingaGalan-Vioque2007,EGUW2019springer} using the \textit{Matlab} package \textit{pde2path} \citep{UeWR2014nmma}. Solution branches are tracked in parameter space via tangential predictors and Newton correctors while varying the main control parameter $\phi_0$.
To take the conservation of both fields into account we impose \eqref{eq:int} as integral side conditions and introduce virtual source terms in \eqref{eq:nondim_bG_final} whose strengths act as further continuation parameters. They are automatically kept at zero during all continuation runs. Following the branches, we numerically calculate the linear stability of the corresponding states and detect bifurcations. This then allows us to switch branches and construct the entire bifurcation diagram. In a few cases, we also continue branches of time-periodic states \citep{Ueck2019ccp}.
In general, we use the $L^2$-norm
\begin{equation}\label{eq:norm}
||\delta \phi|| \equiv \sqrt{\int \sum_{i=1,2}\left(\phi_i - \phi_0\right)^2 {\rm d} x }
\end{equation}
or its temporal average as solution measure.
\section{Linear Stability of homogeneous states} \mylab{sec:linstab}
First, we analyze the linear stability of steady homogeneous states $\vecg{\phi}(x) = (\phi_1(x),\phi_2(x)) = (\phi_0,\phi_0) \equiv \vecg{\phi_0}$ employing the harmonic ansatz
\begin{align}
 \vecg{\phi}(x,t)= \vecg{\phi_0} + \varepsilon \vecg{\widetilde{\phi}} e^{i k x + \lambda t}
 \mylab{eq:ansatz}
\end{align}
Inserting \eqref{eq:ansatz} into Eqs.~\eqref{eq:nondim_bG_final} and linearizing in $\varepsilon \ll 1$ yields
\begin{equation}\label{eq:lambda_original}
\lambda  \vecg{\widetilde{\phi}} = - q^2
\left( \begin{array}{c c}
q^2 + f_1'' & -\left(\rho + \alpha\right) \\
-Q\left(\rho - \alpha\right) & \, Q\left(\kappa q^2 + f_2''\right)
\end{array}\right) \vecg{\widetilde{\phi}} \equiv -q^2\, \tens{B} \,\vecg{\widetilde{\phi}}\,.
\end{equation}
where the second derivatives $f_1''$ and $f_2''$ are evaluated at $\vecg{\phi_0}$, and $q=k/\ell$.
Due to the conserved character of the dynamics any homogeneous state is a steady solution of Eqs.~\eqref{eq:nondim_bG_final} and homogeneous perturbations, i.e.~$k=0$, always correspond to the neutral eigenvalue $\lambda=0$.
Introducing $\tilde\lambda=\lambda/q^2$, and defining the difference in squared coupling strengths $\Delta \equiv \alpha^2 - \rho^2$ gives 
\begin{align}\label{eq:lambda+-}
 \tilde\lambda_\pm = \frac{1}{2} \left[ -\operatorname{tr} \tens{B} \pm \sqrt{(\operatorname{tr} \tens{B})^2 - 4\, \det \tens{B}} \right] \\
  \text{with} \quad \operatorname{tr} \tens{B} = q^2 (1+Q\kappa) + f_1'' + Q f_2'' \quad \text{and} \nonumber\\
  \det \tens{B} = Q\left[q^2 + f_1''\right] \left[\kappa q^2 + f_2''\right] + Q \Delta\,. \nonumber
\end{align}
These rescaled eigenvalues are qualitatively equal to those obtained for generic two-species reaction-diffusion systems as investigated, e.g., by \citet{Turi1952ptrslsbs}.
In consequence, the linear behavior is similar showing a number of different linear instabilities. In addition to the common large-scale stationary (CH) instability, the model also exhibits large-scale oscillatory (Hopf) and small-scale stationary (Turing) instabilities.
For a detailed discussion of equivalences of the present parameters and parameters in the reaction-diffusion system, and an analysis of all occurring instabilities see Sec.~III of \citet{FrWT2021pre}.
Here, we exclusively focus on the pattern forming small-scale stationary instability. Treating $f''_1(\phi_0)$ as primary control parameter, its onset occurs at
\begin{equation}
{f''_1}^\text{T} = \frac1\kappa \left[f''_2 \mp 2 \sqrt{\kappa \Delta} \right]
\mylab{eq:a11tu2}
\end{equation}
with critical wavenumber
\begin{equation}
q_\text{c}^2 =\frac{1}{\kappa}\left[\pm \sqrt{\kappa \Delta} -  f''_2\right]
\mylab{eq:kcTuring}
\end{equation}
where the upper [lower] sign refers to cases $\kappa>1/Q$ [$\kappa<1/Q$].
Applying our specific choices for $f_1''$ and $f_2''$ [Eqs.~\eqref{eq:ff}], choosing $\Delta \phi_0=0$ and using $\phi_0$ as primary control parameter the onset of the Turing instability is at
\begin{equation}
\phi_0^\text{T}= \sqrt{\frac{\pm 2 \sqrt{\kappa  \Delta }-a(1-\kappa)-a_\Delta }{3\left(1- \kappa\right)}}\,.
\mylab{eq:phiT}
\end{equation}
Note that from now on, for simplicity we set $Q=1$.

\begin{figure}[!h]
\centering\includegraphics[width=\textwidth]{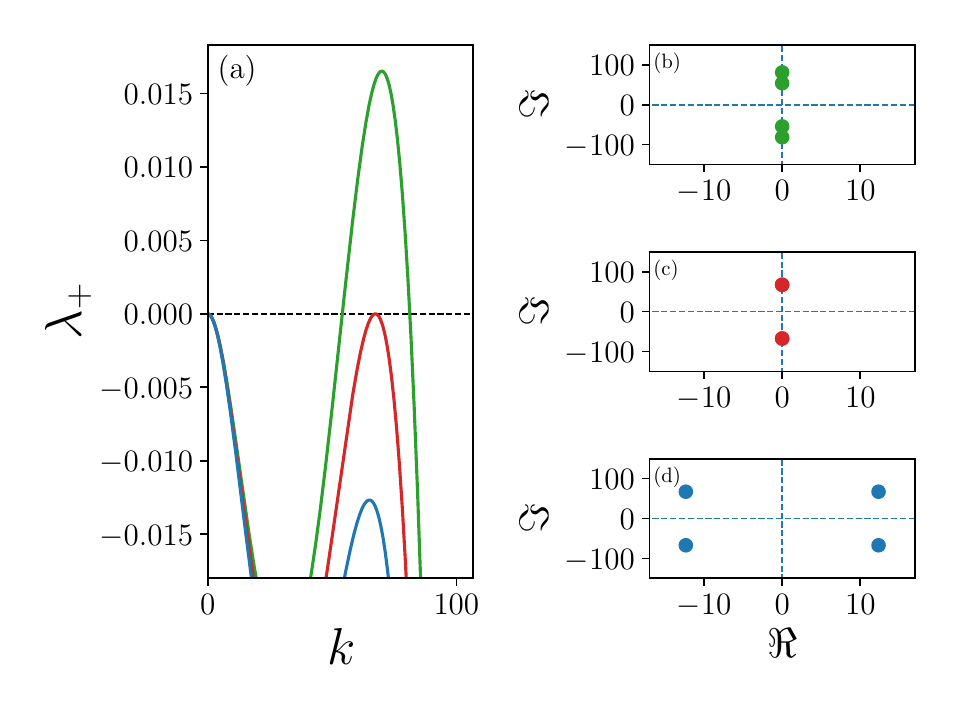}
\caption{The linear behavior near the onset of a small-scale (Turing) instability of the nonvariationally coupled CH equations. Panel (a) pictures three dispersion relations $\lambda_+(k)$: linearly stable case at $|\phi_0|=0.22$ (blue line), neutrally stable case at $|\phi_0|=\phi_0^{\text{T}} \approx 0.211$ [Eq.~\eqref{eq:phiT}, red line] and unstable case at $|\phi_0|= 0.2$  (green line). All eigenvalues are real. Panels (b)-(d) illustrate the corresponding motion of the four spatial eigenvalues in the complex plane at (b) $|\phi_0|=0.2$, (c) $|\phi_0|\approx 0.211$, and (d) at $|\phi_0|=0.22$. The remaining parameters are as in Fig.~\ref{fig:snaking1}.}
\label{fig:dispersion+spatEV} 
\end{figure}
From Eqs.~\eqref{eq:a11tu2} - \eqref{eq:phiT} we see that a Turing instability requires (i) an antisymmetric coupling stronger than the symmetric one, i.e.~$|\alpha|>|\rho|$ implying $\Delta>0$, (ii) $ f''_2 < \pm \sqrt{\kappa \Delta}$ for $\kappa\gtrless1$, (iii) the presence of both rigidities, i.e.~$\kappa\neq 0$ and (iv) unequal rigidities, i.e.~$\kappa \neq 1$. The latter is a direct equivalence of the condition of unequal diffusion constants in the classical reaction-diffusion system \citep{Turi1952ptrslsbs}.
Fig.~\ref{fig:dispersion+spatEV} illustrates the small-scale linear instability showing three dispersion relations for different $|\phi_0|$. For $|\phi_0|>\phi_0^\text{T}$ all eigenvalues are negative (blue line), at the threshold $|\phi_0|=\phi_0^\text{T}$ the maximum touches the $k$-axis at $k=\ell q_\text{c}$ (red line). Decreasing $|\phi_0|$ further, a band of unstable wavenumbers widens around $\ell q_\text{c}$ (green line). In all three cases, all eigenvalues are real and the second branch ($\lambda_-(k)$) is always below zero, hence not shown. 
For different parameters, e.g., for further increasing activity, it is possible that the linear instability threshold approaches the codimension-2 point where the Turing and Hopf instabilities occur simultaneously. Still in the Turing regime, stable modes near $k=0$ may already show complex eigenvalues indicating that oscillatory dynamics may already influence the nonlinear behavior. This hypothesis will be confirmed in Sec.~\ref{sec:snake}.

Next, we consider the spatial eigenvalues of the homogeneous state which give an indication at which parameters localized steady states can be expected. Thus, we consider the time-independent version of Eqs.~\eqref{eq:nondim_bG_final} and apply the ansatz $\vecg{\phi}(x)= \vecg{\phi_0} + \varepsilon \vecg{\widetilde{\phi}} e^{\ell \beta x}$ where $\beta$ is the spatial eigenvalue.
Then the solvability condition, $\det \tens{B}=0$  (with $q^2=-\beta^2$) gives the four spatial eigenvalues
\begin{align}
 \beta= \pm \sqrt{\frac{\kappa f_1'' + f_2'' \pm \sqrt{\left(\kappa f_1'' - f_2''\right)^2 - 4 \kappa \Delta}}{2 \kappa }} \,.
 \mylab{eq:beta}
\end{align}
Their motion in the complex plane is depicted in Fig.~\ref{fig:dispersion+spatEV}~(b)-(d). At $f_1''={f''_1}^\text{T}$ [see panel (c)] they reflect the onset of linear instability, i.e., $\beta=\pm i q_\text{c}$ with double multiplicity.
To study the vicinity of the Turing instability, we specify $f''_1={f''_1}^\text{T} + \varepsilon$  with $|\varepsilon|\ll 1 $ and obtain
\begin{align}
 \beta= & \pm \sqrt{-q_\text{c}^2 \pm \sqrt{\varepsilon} \sqrt{\frac{\varepsilon}{4} +  \mathrm{sgn}(1-\kappa)\sqrt{\frac{\Delta}{\kappa}}} + \frac\varepsilon2}~ \nonumber \\ 
 \approx & \pm \left[i q_\text{c} \pm 
 \frac{\sqrt{\mathrm{sgn}(1-\kappa)}}{2 q_\text{c}} \left(\frac{\Delta}{\kappa}\right)^{\frac14} \sqrt{-\varepsilon} 
\right]
\end{align}
with the sign-function $\mathrm{sgn}(x)$.
If $\kappa>1$ [$\kappa<1$] the homogeneous solution is Turing unstable for $\varepsilon<0$ [$\varepsilon>0$] and the spatial eigenvalues are purely imaginary, see panel (b), as expected for linearly unstable uniform states (note that the real roots of the dispersion relation directly represent the spatial eigenvalues).
  Then in the linearly stable case close to the Turing threshold, i.e., for $\varepsilon>0$ [$\varepsilon<0$], the spatial eigenvalues form a quartet containing one pair of stable and one pair of unstable spatial eigenvalues, i.e., complex conjugated pairs with negative and positive real parts, respectively, because at the Turing threshold the four complex spatial eigenvalues become four imaginary ones.  In consequence, close to the Turing threshold a spatially oscillatory approach to the linearly stable uniform state for $x\to \pm \infty$ is possible. For localized states to exist the spatial eigenvalues should indeed be complex, so the tails of the individual peaks can lock into each other.
Note that further away from the Turing threshold, the spatial eigenvalues can become real which then may indicate the existence of front solutions. The transition from complex conjugated pairs to real spatial eigenvalues occurs at $|\phi_0|$-values also given by Eq.~\eqref{eq:phiT}. In contrast to the Turing instability it demands $f''_2 > \pm \sqrt{\kappa \Delta}$ and, thus, $\beta^2 = -q_c^2 =- \frac{1}{\kappa}\left[\pm \sqrt{\kappa \Delta} -  f''_2\right]$ is positive. 
For the parameters used in Fig.~\ref{fig:snaking1} real spatial eigenvalues appear for $|\phi_0|>0.77$, i.e., far outside the shown range.
In summary, the obtained information on instability type and spatial eigenvalues allows us to conclude that localized states can emerge at $|\phi_0|\approx \phi_0^\text{T}$ under the specified conditions. 
%
\section{Snaking of localized states} \mylab{sec:snake}
%
\begin{figure}[h!]
\includegraphics[width=0.8\textwidth]{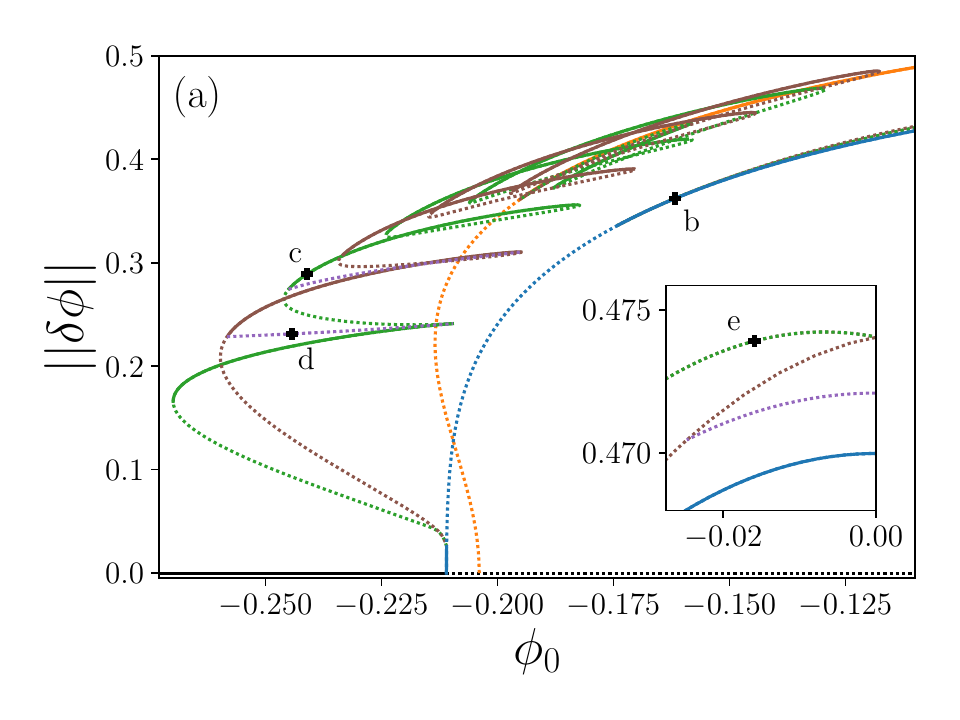}
~\hfill
\includegraphics[width=\textwidth]{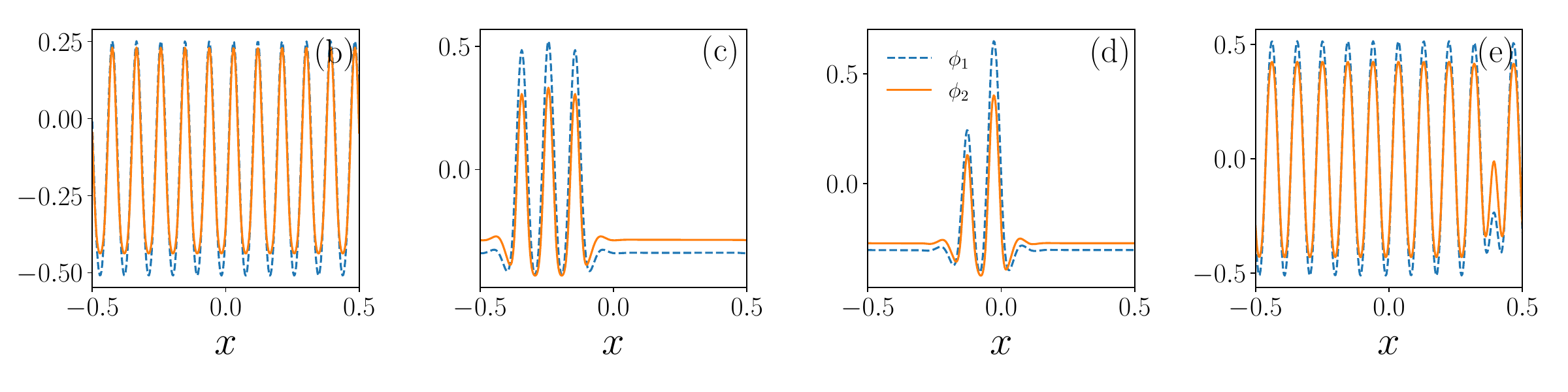}
\caption{\small \it (a) A typical bifurcation diagram for the nonvariationally coupled CH equations is presented focusing on branches of steady localized states. Shown is the $L^2$-norm~\eqref{eq:norm} as a function of the mean concentration $\phi_0$ at fixed $\alpha = 1.65$. Steady symmetric localized states with an even and an odd number of peaks are shown as brown and green lines, respectively. Also included are the two lowest branches of steady asymmetric localized states (purple dashed lines), the periodic steady states with $n=11$ peaks (blue line) and $n=9$ peaks (orange line), and the uniform state (black line). The inset shows branches of localized states near $\phi_0=0$ that emerge at the right most bifurcation of the periodic $n=11$ branch. Solid and dashed lines indicate stable and unstable states, respectively. Panels (b) to (e) show example profiles marked by the corresponding letters in panel (a).
The remaining parameters are $\kappa=0.14$, $a=1.25$, $a_\Delta = -1.9$, $\rho = 1.35$, and $\Delta \phi_0 = 0$. The domain size is $\ell=20 \pi$. }
\label{fig:snaking1} 
\end{figure}

Having identified conditions for a Turing instability, we next investigate the resulting fully nonlinear behavior by addressing the bifurcation behavior of steady states.
We use the common mean concentration $\phi_0$ of the two fields as main control parameter (i.e., always use $\Delta \phi_0= 0$) and discuss the dependency of the bifurcation behavior on the effective temperature $a$ and on the strength of antisymmetric coupling $\alpha$. Our present choice $\Delta \phi_0= 0$ results in the inversion symmetry $(\phi_i,\phi_0)\to (-\phi_i,-\phi_0)$. Hence, all bifurcation diagrams obtained here for $\phi_0<0$ can be reflected at $\phi_0=0$ to obtain the complete picture.
Note that all numeric results are obtained for a finite domain employing periodic boundary conditions. This implies that all our ``localized states'' are not truly localized states in the asymptotic sense but their numeric finite-domain approximations. However, these states converge to truly localized states when the domain size approaches infinity. This approach is often followed to numerically obtain bifurcation diagrams featuring branches of localized states (e.g.~\citet{BuKn2006pre,TARG2013pre}).

Fig.~\ref{fig:snaking1}~(a) gives a typical bifurcation diagram and provides the main message of this work. It illustrates that the model features families of localized states that form the slanted (or tilted) snakes-and-ladders structure typical for homoclinic snaking in systems with a conservation law \citep{Knob2016jam}. At the chosen parameter values ($a=1.25$ and $\alpha=1.65$), the uniform state is linearly stable for $|\phi_0|>\phi_0^\text{T} \approx 0.211$ [see Eq.~\eqref{eq:phiT}]. At $\phi_0=-\phi_0^\text{T}$ the first primary bifurcation occurs where a branch of periodic steady states with 11 peaks emerges supercritically [blue line, cf.~profile in Fig.~\ref{fig:snaking1}~(b). We call it ``periodic $n=11$ branch'']. We also show the periodic $n=9$ branch that bifurcates subcritically at slightly smaller $|\phi_0|$ (orange line). Further branches of $n$-peak periodic states emerge at $|\phi_0|<\phi_0^\text{T}$ (not shown).

The periodic $n=11$ branch is first stable but soon loses its stability in a secondary pitchfork bifurcation where two branches of steady symmetric localized states emerge subcritically.
Note that the locus of this secondary bifurcation depends on the domain size. For an infinite domain, the branches of localized states bifurcate together with the branch of periodic states from the homogeneous solution at the first primary bifurcation \citep{BBKM2008pre}. The emergence of localized states even in the case of a supercritically emerging branch of periodic states is possible because of the conservation law(s). It allows for the so-called Matthews-Cox instability \citep{MaCo2000non} related to the existence of the resulting neutral mode at $k=0$. The Matthews-Cox instability is an extension of the well known Eckhaus instability responsible for secondary bifurcations where side-band solutions gain stability in systems without conservation laws \citep{TuBa1990pd}.
  
Here, the two branches emerging at the secondary bifurcation consist of localized states with an even (brown line) and odd number [green line, cf.~profile in Fig.~\ref{fig:snaking1}~(c)] of peaks, respectively. The branches first continue towards smaller $\phi_0$ before they turn in respective saddle-node bifurcations. Further saddle-node bifurcations follow, give the branches a snaking appearance, and are related to the addition of pairs of peaks to the localized state. Note that part of the tilted snaking structure lies in the Turing unstable region, i.e., has $\phi_0<-\phi_0^\text{T}$. This, however, does not contradict our previous argument based on the spatial eigenvalues of the uniform background state as for the considered finite domain and the conservation law, the $\phi$-value of the background  is in the presence of a localized patterned patch
  smaller than $\phi_0$. With other words, at coexistence of a uniform state (the background) and a periodic pattern (the localized patch) the local mean concentrations in the two regions differ allowing for a Turing-stable background at Turing-unstable overall mean concentration.

Following the snaking branches, the localized patterned patch increases in extension. When the entire domain is filled, both branches terminate subcritically on the periodic $n=9$ branch, and the latter gains stability. Whether the branches of localized states end on the same branch of periodic states they emerged from depends on the used parameters, in particular, on the domain size. It is also possible that branches of even and odd localized states terminate on different period-$n$ states. For further details on this finite size effect see \citet{BBKM2008pre}. 

Furthermore, the slanted snakes-and-ladders structure contains the usual branches of unstable asymmetric localized states (purple lines) that connect the two branches of symmetric localized states via pitchfork bifurcations located in the vicinity of their saddle-node bifurcations. An example profile with two peaks from the first rung is given in Fig.~\ref{fig:snaking1}~(d). Note that all of these states are steady. In case of the asymmetric states this is remarkable for the present nonvariational model and will be further discussed elsewhere.

Note that the periodic $n=11$ branch also shows a stabilising pitchfork bifurcation at $\phi_0\approx-0.17$. There, two further branches of localized states emerge and proceed towards larger $\phi_0$ [in Fig.~\ref{fig:snaking1}~(a) they visually nearly coincide with the branch of periodic states]. After passing $\phi_0=0$ they 
terminate in the mirrored bifurcation at positive $\phi_0$. States on these odd and even branches show a localized defect structure corresponding to a respective 1-peak or 2-peak hole within the periodic pattern [see, e.g.,  the profile in Fig.~\ref{fig:snaking1}~(e)]. These branches also show a (very reduced) snaking behavior and exchange stability via an asymmetric rung state (purple line in inset). However, we found them to always be unstable.

\begin{figure}
\centering
\includegraphics[width=0.75\textwidth]{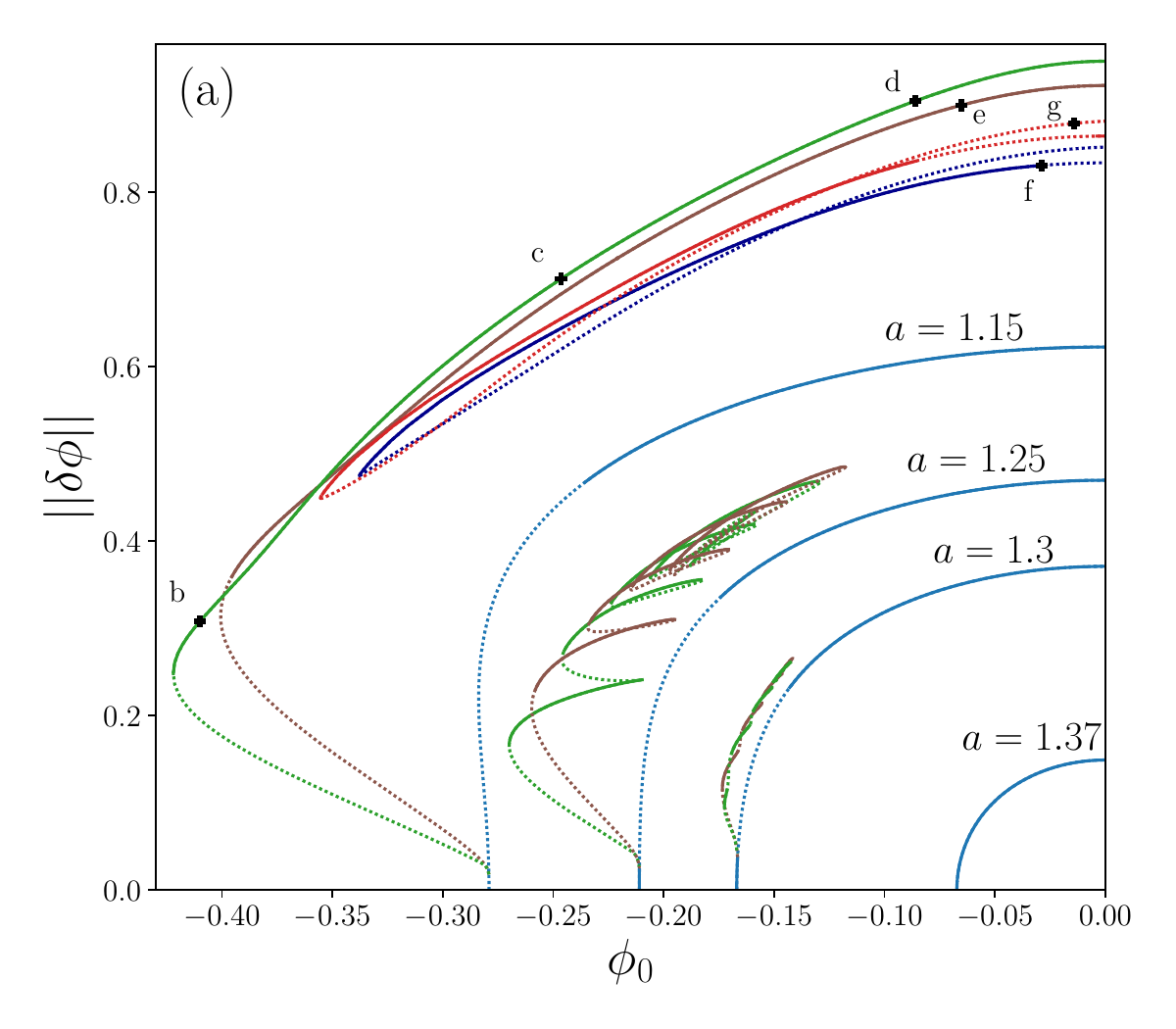}~\\
\includegraphics[width=0.8\textwidth]{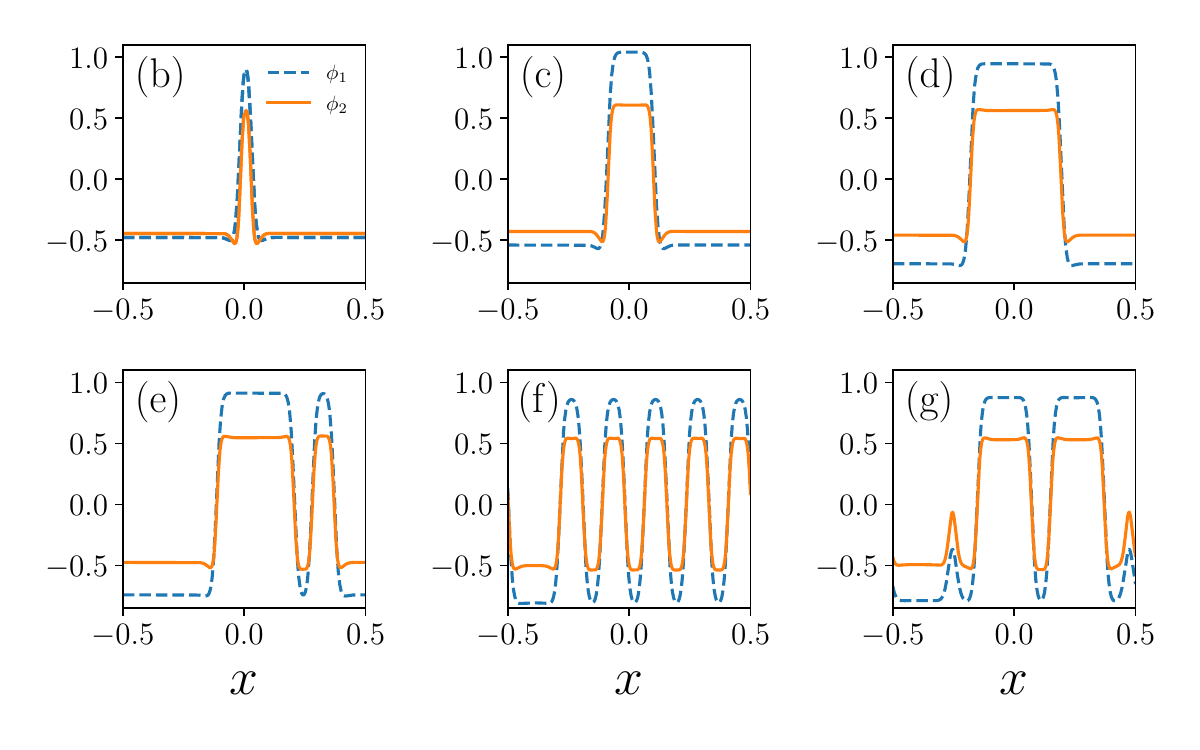}
\caption{\small \it (a) The dependency of the entire slanted snakes-and-ladders structure on the effective temperature is shown. From left to right, the bifurcation structure as a function of $\phi_0$ is shown for $a=1.15$, $1.25$, $1.3$ and $1.37$. Panels (b) to (g) present concentration profiles at accordingly marked loci in panel (a). Dark blue and red lines correspond to two isolas for $a=1.15$. The remaining line styles and parameters are as in Fig.~\ref{fig:snaking1}.}
\label{fig:snaking_a} 
\end{figure}

Next, we investigate the changes in the slanted snakes-and-ladders structure that occur when the effective temperature $a$ is varied. Fig.~\ref{fig:snaking_a}~(a) compares the behavior at four different values of $a$. The second lowest value, $a=1.25$, corresponds to the structure known from Fig.~\ref{fig:snaking1}~(a). In each case we show the periodic $n=11$ branch [blue lines] and the branches of odd [green lines] and even [brown lines] localized states. Increasing $a$ shifts the entire structure toward larger $\phi_0$, while neighboring saddle-node bifurcations on the snaking branches approach each other, thereby narrowing the snaking structure. Eventually, the saddle-node bifurcations pairwise annihilate in subsequent hysteresis bifurcations. At $a=1.3$ most of the saddle-node bifurcations still exist, only the ones at the largest peak numbers have already annihilated. Upon further increase in $a$, first all saddle-node bifurcations vanish, then both secondary bifurcations where the localized states emerge become supercritical. Finally, they approach each other and annihilate. This is all similar to behavior seen for phase-field-crystal models \citep{TARG2013pre,HAGK2020arxiv}. It is notable that even without saddle-node bifurcations the two branches of localized states still wiggle about each other and exchange stabilities via the rung branches of asymmetric states that connect them. The resulting monotonic structure is termed smooth snaking \citep{DaLi2010sjads}. The final example in Fig.~\ref{fig:snaking_a}~(a) at $a=1.37$ is beyond the point where all localized states connected to the $n=11$ branch have disappeared and only the branch of stable periodic states remains.

Furthermore, for sufficiently increased $a$, the snaking structure tends to connect at both ends to the same periodic $n$-peak branch. E.g., while at $a=1.25$ the branches of localized states emerge on the periodic $n=11$ branch and terminate on the $n=9$ branch, for  $a=1.3$ they end on the $n=10$ branch (not shown), and upon further increase in $a$ on the $n=11$ branch (not shown). This behavior is natural since for a narrowing snaking structure in the bifurcation diagram the corresponding states loose their strongly localized character, i.e., the homogeneous part becomes spatially more modified and the localized structure resembles a wave packet. Finally, the localized states correspond to spatially modulated periodic states with a constant number of peaks, in our case 11. Thus the branches necessarily end on the $n=11$ branch. The transitions are related to reconnection between different snaking structures (not shown). 

Considering again our $a=1.25$ reference case of Fig.~\ref{fig:snaking1}, we next discuss a decrease in $a$. Then, a transition from a supercritical to a subcritical primary bifurcation occurs at $a\approx 1.23$.  Already at $a=1.15$, the bifurcation structure has dramatically changed [Fig.~\ref{fig:snaking_a}~(a)]: As the width of the snaking region strongly increases with decreasing $a$, eventually, the rightmost saddle-node bifurcations of the snaking structure at negative $\phi_0$ collide with their images from the mirroring snaking at positive $\phi_0$. Other structures of localized states like the hole states mentioned above may also be involved. Upon collision, saddle-node bifurcations annihilate and the formerly continuous snakes-and-ladders structure successively breaks up into an increasing number of disconnected pieces, mostly in the form of isolas, see e.g., the dark blue and the red line in Fig.~\ref{fig:snaking_a}~(a). Only the two branches of one-peak (green line) and two-peak (brown line) localized states are still connected to the branch of periodic states. 
Thereby, when following the one-peak branch towards larger norms, the localized peak widens [see Figs.~\ref{fig:snaking_a}~(b) to (d)] and forms a plateau-like structure. 
  In the context of a Swift-Hohenberg equation one would call these states top-hat states, but in the present case of coupled CH equations one could better speak of a fully phase-separated state in analogy to the classical CH case. However, in contrast to the classical variational case, where any pattern would coarsen into the phase-separated state, here, other linearly stable states exist. These are, in particular, other patterns combining different peaks and/or plateaus. Fig.~\ref{fig:snaking_a}~(e) shows a two-peak state which combines a narrow peak and a broader plateau. In addition, Fig.~\ref{fig:snaking_a}~(a) includes two isola structures for $a=1.15$ that consist of localized states of four [red line in (a), profile in panel (g)] and five structures [dark blue line in (a), cf.~profile in panel (f)], respectively. Their left folds correspond to the second and third left hand side fold of the branch of localized states of even and odd peak number at $a=1.25$, respectively. There exist further isola-like branches of localized states at $a=1.15$ that are not included in the figure. Note that for all mentioned plateau states the approach to the plateau concentrations is oscillatory and not monotonic as in the classical CH case.

\begin{figure}[h!]
\centering
\includegraphics[width=0.8\textwidth]{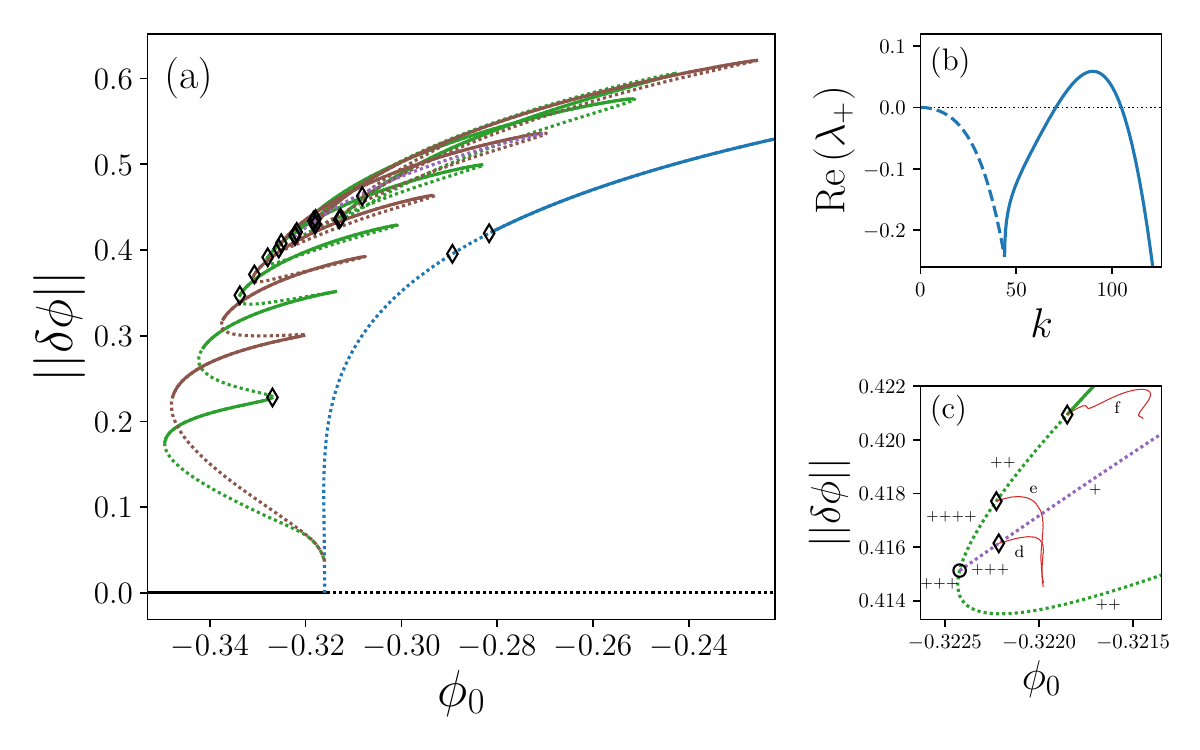}~\\
\vspace*{-0.2cm}
\includegraphics[width=\textwidth]{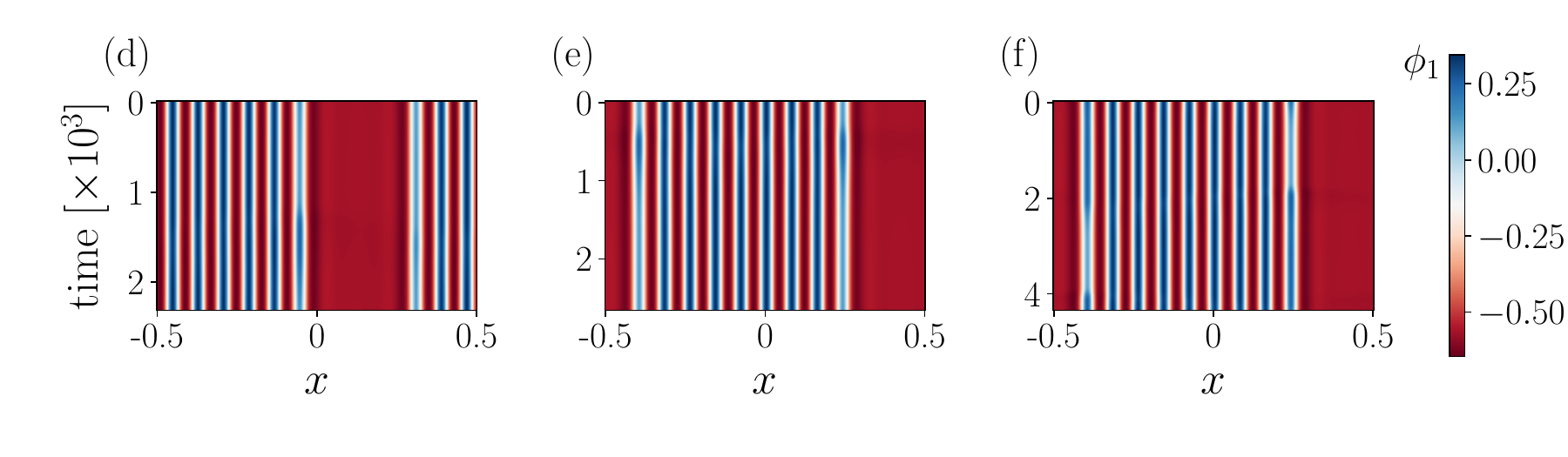}
\caption{\small \it (a) Bifurcation diagram in dependence of $\phi_0$ illustrating the appearance of time-dependent behavior within the slanted snakes-and-ladders structure at activity $\alpha = 1.78$, effective temperature $a=0.9$ and reciprocal coupling $\rho=1.35$. The diamond symbols represent Hopf bifurcations where branches of time-periodic states emerge. The remaining line styles and parameters are as in Fig.~\ref{fig:snaking1}.
Panel (b) gives a dispersion relation at $\phi_0=-0.3$ where solid [dashed] lines indicate real [complex] eigenvalues. Panel (c) magnifies a region of (a) near the emergence of the 8th rung state where Hopf bifurcations occur on branches of steady symmetric and asymmetric localized states. The number of unstable eigenvalues are indicated by ``+''-symbols. Branches of time-periodic states are shown as thin red solid lines. The space-time plots in panels (d) to (f) show a state on each of the branches of time-periodic states marked ``d'' to ``f'' in (c), respectively.
}
\label{fig:snaking_hopf} 
\end{figure}

Finally, we discuss qualitative changes in the bifurcation structure that occur when the nonreciprocal coupling strength $|\alpha|$ is varied. 
Proceeding as before and taking Fig.~\ref{fig:snaking1} as reference case, we observe a very similar behavior when changing activity $\alpha$ instead of temperature $a$. Namely,
  for an increase  of  $\alpha$ at fixed $a=1.25$ we find a transition from slanted snaking with saddle-node bifurcations via smooth slanted snaking to complete suppression of localized states (all already below $\alpha=1.7$), similar to the transition that occurs in Fig.~\ref{fig:snaking_a} when $a$ is increased first to $a = 1.3$ and then to $a = 1.37$ at fixed $\alpha = 1.65$. Decreasing $\alpha$, e.g., to $\alpha = 1.6$, the snaking structure breaks up into separated parts via necking bifurcations involving the mirroring snaking structure at positive $\phi_0$. That is, isolas form similar to the case $a = 1.15$, $\alpha = 1.65$ in Fig.~\ref{fig:snaking_a}. Due to the strong similarities we do not include the bifurcation diagrams.
However, a difference between the influence of $a$ and $\alpha$ is that the former does not affect the critical wavenumber $q_c$ [insert Eq.~\eqref{eq:phiT} into Eq.~\eqref{eq:kcTuring}], i.e., temperature does not affect the transition from large-scale (CH) to small-scale (Turing) instability. Instead it only influences the value $\phi_0^T$ [Eq.~\eqref{eq:phiT}] at onset. In contrast, the activity controls the transition between the two instability types. For decreasing $\alpha$, e.g., at $\alpha=1.5$, the system exhibits a large-scale instability. Then the branch of one-peak states of the original localized snaking structure subcritically emerges from the primary bifurcation of the homogeneous branch instead from a secondary bifurcation as for $a=1.15$ and $\alpha=1.65$ in Fig.~\ref{fig:snaking_a}~(a).
Another major difference between $a$ and $\alpha$ is that the latter is responsible for the occurrence of time-periodic behavior, as e.g., already noted above when discussing the linear analysis. How this affects the snaking structure is discussed next.

\begin{figure}[h]
\centering
\includegraphics[width=0.95\textwidth]{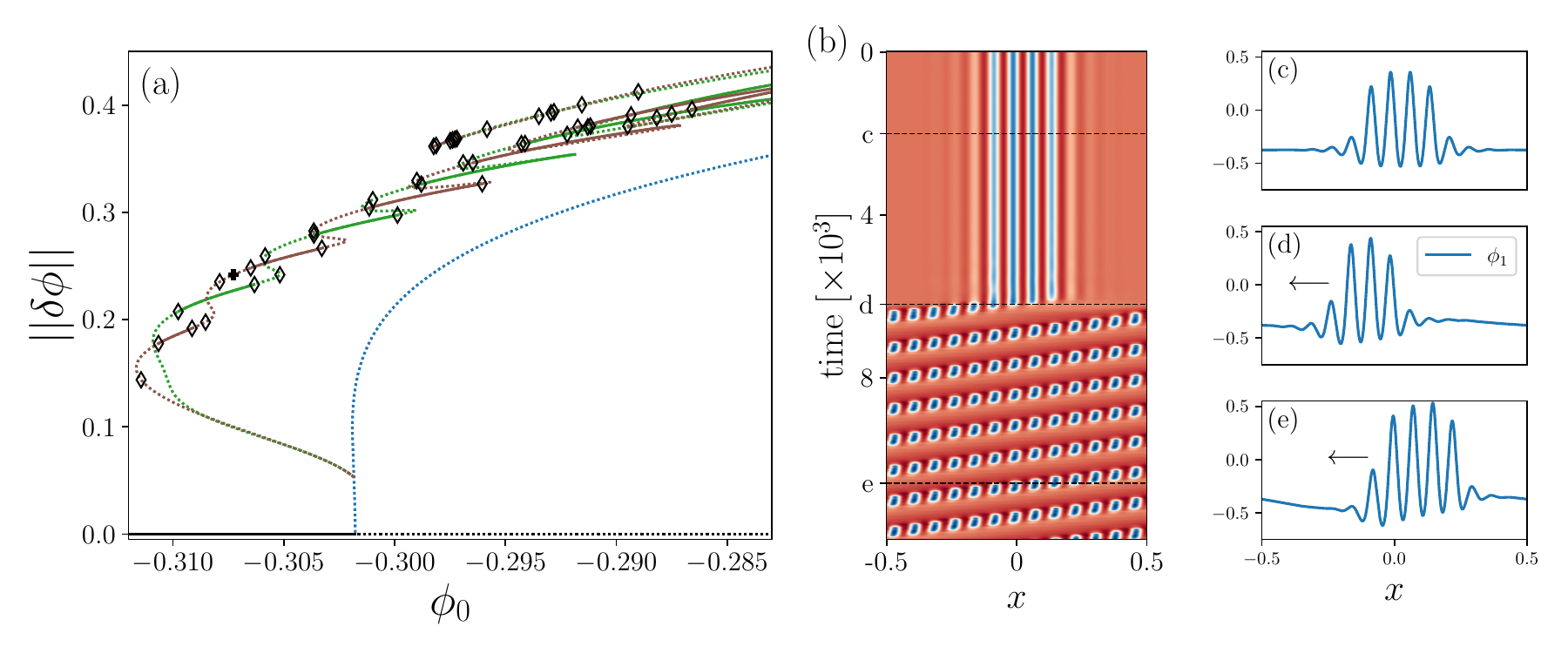}
\caption{\small \it (a) Bifurcation diagram in dependence of $\phi_0$ at moderately large activity $\alpha = 1.8$. It illustrates the appearance of many more Hopf bifurcations (diamond symbols) within the slanted snakes-and-ladders structure as compared to Fig.~\ref{fig:snaking_hopf}. The remaining parameters and line styles are as in Fig.~\ref{fig:snaking_hopf}. Panel (b) is a space-time plot of $\phi_1$ and illustrates the emergence of a drifting localized state with temporally oscillating peaks. The initial state is the even, linearly unstable localized state marked by a ``+'' in panel (a). Panels (c) to (e) give three snapshots at times marked by thin dotted lines in (b). Drift direction is indicated by an arrow.
}
\label{fig:snaking_hopf2} 
\end{figure}

Using the example in Fig.~\ref{fig:snaking1} as reference case, we keep the symmetric coupling at $\rho = 1.35$ and increase $\alpha$. We also adjust $a$ to stay in the parameter region of slanted snaking. The resulting bifurcation diagram for $\alpha=1.78$ and $a=0.9$ is presented in Fig.~\ref{fig:snaking_hopf}~(a) while panel (b) gives a corresponding dispersion relation for the homogeneous state at $\phi_0=-0.3$: A band of stable oscillatory modes exists at wavenumbers to the left of the band of unstable stationary modes. Although, there is no direct influence of the oscillatory modes on the linear behavior, we expect an impact on the fully nonlinear behavior.

Indeed, at $\alpha=1.78$ a dramatic change has occurred: On the slanted homoclinic snaking structure of steady localized states with the usual sequences of saddle-node and pitchfork bifurcations, additionally, a number of Hopf bifurcations has appeared (marked by diamond symbols). They are found on the branches of steady symmetric and asymmetric localized states as well as on the branch of periodic steady states. One notes that the Hopf bifurcations predominantly occur near the left hand side saddle-node bifurcations of the snaking branches. Only one of them is found close to the right hand side, namely, the one near the lowest saddle-node bifurcation.  On the left hand side, more and more such Hopf bifurcations occur the more extended the localized structure becomes. In particular, none exists near the four lowest left hand side saddle-node bifurcations, whereas one Hopf bifurcation is observed near each of the next three. Two such bifurcations appear near each of the four uppermost left saddle-node bifurcations. One example of the latter behavior is highlighted in Fig.~\ref{fig:snaking_hopf}~(c), which magnifies the region of panel~(a) where the 8th rung state [purple line in (a) and (c)] emerges.

In Fig.~\ref{fig:snaking_hopf}~(c), the ``+''-symbols represent the unstable eigenvalues and indicate how the appearance of Hopf bifurcations affects the character of saddle-node and pitchfork (circle symbol) bifurcation. Without Hopf bifurcations, the ascending snaking branch would first pass from two to one unstable eigenvalues at the saddle-node bifurcation, and then gain stability at the subcritical pitchfork bifurcation where the asymmetric rung state emerges. Now, however, at sufficiently large activity, saddle-node and (now supercritical) pitchfork bifurcation both further destabilize the branch such that it carries four unstable eigenvalues. These are subsequently stabilized via two Hopf bifurcations. Furthermore, the rung state emerges with three unstable eigenvalues, two of which are stabilized in another Hopf bifurcation.

Numerically continuing the emerging branches of time-periodic states is challenging since their time periods are large and even diverge when continuing along the branches. Therefore, with our  present  numerical means we are limited to part of the branches. Nevertheless, the obtained behavior gives a consistent picture:  The three calculated branches are shown in panel~(c) as thin red solid lines. They represent standing waves [cf.~space-time plots of field $\phi_1$ in Figs.~\ref{fig:snaking_hopf}~(d)-(f)] and emerge subcritically, hence, are unstable. Closely inspecting the space-time plots one notices that although the central peaks of the localized states barely change in time, the outermost peaks strongly oscillate - almost vanishing and reappearing in the course of a temporal period. Note that $\phi_2$ behaves similarly.

First, we consider the two branches which emerge from the symmetric steady states (green line). On the left hand branch [Fig.~\ref{fig:snaking_hopf}~(e)] we find standing waves that remain spatially symmetric at all times, i.e., the outer peaks oscillate in phase. In contrast, the standing waves on the right hand branch [Fig.~\ref{fig:snaking_hopf}~(f)] shows a broken reflection symmetry, i.e., the outer peaks oscillate in anti-phase (but keeps with respect to a combined spatial reflection and temporal shift by half a period). Corresponding to their symmetries, the former  and latter branch then approach a homoclinic bifurcation on the lower (more unstable) part of the branch of symmetric steady states and in a heteroclinic bifurcation on the branch of unstable asymmetric steady states, respectively. Note that the latter branch exists twice, with the associated states related via reflection symmetry. The termination in a global bifurcation is indicated by the seemingly diverging time period. It is observed that both Hopf bifurcations emerge in Bogdanov-Takens bifurcations at the saddle-node of the steady state. 
At these codimension-2 bifurcations, i.e. at a critical value of a second control parameter \cite{Strogatz2014}, here the activity, a Hopf bifurcation and a homoclinic bifurcation are created together with the connecting branch of time-periodic states. Increasing the activity beyond such a critical value the created bifurcations move apart (on different branches of steady states) and the branch of time-periodic states becomes longer. Hence at begin both Hopf bifurcations are connected to the unstable part of the symmetric localized steady state. For discussions of such scenario in other nonvariational CH type equations see, e.g., \citet{KoTh2014n,TALT2020n}.

For the asymmetric time-periodic states [Fig.~\ref{fig:snaking_hopf}~(d)] that emerge at the Hopf bifurcation on the branch of asymmetric steady states only one of the two outer peaks oscillates strongly while the other remains nearly unchanged. In part of the period the state is nearly symmetric. As this part increasingly dominates as one follows the branch, it also terminates on the branch of spatially symmetric steady states. It is observed that this Hopf bifurcation emerges when the right most Hopf bifurcation on the symmetric branch crosses the pitchfork bifurcation.  Therefore, the original Hopf bifurcation splits into two, one moving along the asymmetric steady state branch and connected to the symmetric steady state and the other moving further along the symmetric branch but now connected to the asymmetric one.

The branches of time-periodic states which emerge in the two Hopf bifurcations on the branch of the periodic steady states (blue line) seem to terminate in homoclinic bifurcations on the unstable part of the same branch and hardly show any time dependence (not shown).

The appearance of Hopf bifurcations extends to the entire snakes-and-ladders structure when the activity is further increased. Thus, for a small increase to $\alpha=1.8$, Fig.~\ref{fig:snaking_hopf2} shows the appearance of one or more Hopf bifurcations near almost all saddle-node bifurcations. While the activity increases all Hopf bifurcations migrate towards the stable parts of the snaking branches, causing them to shrink until the steady localized states are completely destabilized when Hopf bifurcations from the left and right collide.
Along with this, stable time-periodic localized structures establish as shown by the space-time plot of field $\phi_1$ in Fig.~\ref{fig:snaking_hopf2}~(b) and the selected snapshots in panels (c)-(e). Again, field $\phi_2$ is not shown, but behaves similarly. In panel (b) the initial condition is the 4 peak state, marked by a bold ``+''-symbol in panel (a), which carries two unstable eigenvalues due to the Hopf bifurcation where the localized steady states get unstable for decreasing $\phi_0$. For an initial time this state remains almost constant [cf. panel (c)] until at $t\approx 5.6 \cdot 10^3$ the outer two peaks begin to oscillate in anti phase which represents the time periodic behavior also encountered in Fig.~\ref{fig:snaking_hopf}~(f). The oscillating state, however, turns out to be unstable as well, and at $t\approx 6.2 \cdot 10^3$ the whole structure stays asymmetric with a more dominant left outer peak and starts moving to the left [cf. panel (d)]. While the asymmetry becomes more pronounced, the velocity increases and the peaks develop a higher amplitude. Finally, for $t\gtrapprox 10 \cdot 10^3$ an asymmetric localized state is established, which moves with constant drift velocity while its five peaks oscillate continuously. Similar stable time periodic localized states with different number of peaks are also found at other parts of the snakes-and-ladders structure where the usually stable steady localized states are destabilized via Hopf bifurcations. Due to the superposition of oscillation and drift, we expect these states to arise in drift-pitchfork bifurcations from standing wave branches that are similar to those shown in Fig.~\ref{fig:snaking_hopf}~(c).

Note finally that at the considered parameters, there exist no drift-pitchfork bifurcations on the steady localized state branches, that are related to the emergence of traveling localized states of constant shape in other active media models like, e.g., the active phase-field-crystal model \citep{OpGT2018pre}. Following the approach of \citet{OpGT2018pre} also here one may derive a condition, $0= \int \phi_1^2 + \frac{\rho + \alpha}{\rho -\alpha}\, \phi_2^2 \, {\rm d}x$, that indicates where drift bifurcations occur. However, for the shown branches of steady states, this condition is nowhere fulfilled.

\section{Conclusion}
\mylab{sec:conc}

We have explored a CH model consisting of two conservation laws with linear nonvariational (nonreciprocal) coupling. In contrast to expectations for CH models we have shown that above a critical activity the system can show a Turing instability and also exhibits slanted homoclinic snaking of localized structures, i.e., branches of symmetric and asymmetric localized states form a snakes-and-ladders structure.

Such features do not exist for the standard passive CH models. They describe decomposition into phase-separated structures that continuously coarse to approach thermodynamic equilibrium. Hence, it is the linear nonvariational coupling which dramatically changes the systems behavior. These findings are very much in line with recent studies of the intriguing impact of nonvariational interactions between order parameter fields, see e.g., \citep{MeLo2013prl,SATB2014c,OpGT2018pre,SaAG2020,YoBM2020pnas,TSJT2020pre}. 

In particular, here, localized states do only exist if the system is active, i.e., if the nonreciprocal coupling dominates the reciprocal one. Then, in general, activity can induce a small-scale instability in a CH system, normally characterized by large-scale instabilities. In consequence, stable periodic states are possible and coarsening can be entirely suppressed, as further investigated in \citet{FrWT2021pre}. Coarsening and its arrest in such a system is also investigated in \citet{SaAG2020} by direct time simulations.

The emerging snakes-and-ladders structure is very similar to the one found in standard pattern forming systems with a conservation law \citep{Knob2016jam}, e.g., in the conserved Swift-Hohenberg (or phase-field-crystal) model \citep{TARG2013pre} and its active variants \citep{OpGT2018pre,OKGT2020c,HAGK2020arxiv}. Correspondingly, we observe a similar change in behavior for increasing effective temperature $a$. Namely, we see a transition from ``broken snaking'' featuring disconnected isolas, to well developed slanted snaking showing the snakes-and-ladders structure with saddle-node bifurcations, to smooth slanted snaking without saddle-node bifurcations, to the absence of localized states - the sequence seems to be generic for localized states of conserved quantities and can also be seen when increasing activity instead of temperature.

However, in contrast to the variational case of a gradient dynamics system where slanted snaking with the mean concentration as control parameter may be mapped to vertically aligned snaking when the chemical potential is employed as control parameter \citep{TARG2013pre,HAGK2020arxiv}, for the present active system no such direct mapping can be expected: Although one could define nonequilibrium chemical potentials there is no Maxwell line about which the snaking could be centered. However, this could be further investigated in the future together with a direct comparison of the present system to its counterpart with nonconserved dynamics, namely, nonvariationally coupled Allen-Cahn equations.

Another crucial difference to passive models is the existence of branches of time-periodic states that appear at moderately large activity. There, a band of oscillatory instability modes appears in the dispersion relation for the uniform state as well as many Hopf bifurcations within the snakes-and-ladders structure. However, the relation between the two observations needs further investigation. In consequence of the emergence of the Hopf bifurcations (normally, at Bogdanov-Takens bifurcations), the parameter regions where steady linearly stable localized states dominate dramatically shrinks. In turn, time simulations converge to stable time-periodic localized states.

Oscillatory instabilities within the snakes-and-ladders structure are also reported in nonvariational Swift-Hohenberg models \citep{BuDa2012sjads}, there, however, Hopf bifurcations on the asymmetric rung states seem to be absent. The major difference between the spatio-temporal patterns in our case and the ones observed in the nonvariational Swift-Hohenberg model of \citet{BuDa2012sjads} and in the conceptually related nonvariational CH model \citep{SATB2014c} is the conservation property. In the two mentioned studies the appearing spatio-temporal patterns are not restricted by mass conservation, hence, concentrations can vary during time and along any solution branch which makes them qualitatively different to the ones presented here where both concentrations are always preserved. 

In our study we have limited our investigation of localized states to the case $\alpha > \rho$. First results show that localized states also appear for $\alpha<- \rho$. This is not surprising since the linear behavior is invariant w.r.t.~the sign of $\alpha$, hence, the small-scale instability necessarily occurs. In contrast to the case described here, the fields are in anti-phase, i.e. peaks of $\phi_1$ pin to holes of $\phi_2$ and vice versa. These localized states can still exhibit snakes-and-ladders structures, but, in particular regarding time-periodic states, one can expect a different behavior. This should be further investigated in future work.

Note, finally, that the asymmetric localized states that form the rung states of the snakes-and-ladders structure remain at rest even at strong activity. This is somewhat unexpected as asymmetric states in nonvariational models are normally expected to drift, see e.g., \citet{HoKn2011pre,BuDa2012sjads}. However, such states are also found in a number of active PFC models \citep{OpGT2018pre,HAGK2020arxiv}. This will be further analyzed elsewhere.

\vskip2pc

\noindent

\section*{Acknowledgment}

We acknowledge frequent discussions with S.V. Gurevich, M.P. Holl and E. Knobloch, in particular, about impossible steady asymmetric states. The work is supported by the doctoral school ‘‘Active living fluids’’ funded by the German French University (Grant No. CDFA-01-14). 
T.F.H. thanks the foundation ``Studienstiftung des deutschen Volkes'' for financial support.




\end{document}